\documentclass[showpacs,aps,prd,nofootinbib,floatfix,amsmath,amssymb]{revtex4}
\usepackage{graphicx}
\begin{document}

%\tightenlines

\title{$SU_L(2)\times U_Y(1)$-invariant description of the bilepton contribution
to the $WWV$ vertex in the minimal 331 model}
\author{J. Monta\~no}
\email[E-mail:]{montano@fcfm.buap.mx}
\author{G. Tavares-Velasco}
\email[E-mail:]{gtv@fcfm.buap.mx}
\author{J. J. Toscano}
\email[E-mail:]{jtoscano@fcfm.buap.mx} \affiliation{Facultad de
Ciencias F\'{\i}sico Matem\'aticas, Benem\'erita Universidad
Aut\'onoma de Puebla, Apartado Postal 1152, Puebla, Pue., M\'exico}
\author{F. Ram\'\i rez-Zavaleta}
\email[E-mail:]{rzf@fis.cinvestav.mx} \affiliation{Departamento de
F\'{\i}sica, CINVESTAV, Apartado Postal 14-740, 07000, M\'exico D.
F., M\'exico}

\date{\today}

\begin{abstract}
We study the one-loop sensitivity of the $WWV$ ($V=\gamma,\,Z$)
vertex to the new massive gauge bosons predicted by the minimal
$SU_L(3)\times U_X(1)$ model, which have unusual couplings to the
standard model (SM) gauge bosons. A gauge-fixing procedure covariant
under the $SU_L(2)\times U_Y(1)$  group was introduced for these new
gauge bosons (dubbed bileptons) in order to generate gauge-invariant
Green functions. The similarities between this procedure and the
nonconventional quantization scheme of the background field method
are discussed. It is found that, for relatively light bileptons,
with a mass ranging from $2m_W$ to $6m_W$, the radiative corrections
to the form factors associated with the $WWV$ vertex can be of the
same order of magnitude than the SM one. In the case of heavier
bileptons, their contribution is smaller by about one and two orders
of magnitude than their SM counterpart.
\end{abstract}


\maketitle

\section{Introduction}

The  $WWV$ ($V=\gamma, Z$) one-loop structure has been considerably
studied in the literature not just because it may constitute a
mechanism through which physics beyond the Fermi scale may show up,
but also due to some theoretical issues concerning its dependence on
the gauge-fixing scheme. It turns out that the conventional
gauge-fixing procedures give rise to ill-behaved off-shell Green
functions that may display inadequate properties such as a
nontrivial dependence on the gauge-fixing parameter, an increase
larger than the one observed in physical amplitudes at high
energies, and the appearance of unphysical thresholds. The on-shell
Green functions can represent physical amplitudes as they are
independent on the gauge-fixing procedure, such as occurs with the
static electromagnetic properties of the $W$ boson
\cite{Bardin,STATICPW,WWg331}, but gauge independence is lost if at
least one external particle  becomes virtual. Although off-shell
Green functions are generally gauge dependent, the $S$-matrix
elements to which they contribute must be gauge independent. This is
the case of the off-shell $WWV$ vertex, which is just a piece of
some physical process such as the $e^+e^-\to W^+W^-$ and $\gamma
\gamma \to W^+W^-$ reactions. Nonetheless, it would be interesting
if one was able to study the sensitivity to radiative corrections of
the $WWV$ coupling, and other SM couplings as well, without invoking
some particular $S$-matrix element.

The concepts of gauge invariance and gauge independence are two
essential ingredients of gauge systems, though the former is not
necessarily present at the quantum level. While gauge invariance
plays a central role when defining the classical action of the
system, once the latter is quantized one must invariably invoke an
appropriate gauge-fixing procedure to define a nondegenerate action,
which means that gauge invariance is to be broken explicitly. The
resultant action is not gauge invariant, though it is invariant
under BRST symmetry \cite{BRST}. As a consequence, the Green
functions derived from this action cannot satisfy simple (QED-like)
Ward identities, but they do satisfy more elaborate Slavnov-Taylor
identities that are dictated by BRST symmetry. Also, Green functions
contain much unphysical information that is removed provided a
physical observable is considered. Contrary to Green functions,
which are highly dependent on the gauge-fixing procedure, physical
amplitudes have no such dependence, thereby being gauge independent.
There are thus some subtle mechanisms that conspire to produce
nontrivial cancelations between the Green functions defining a
physical observable. It is clear that a nonconventional quantization
scheme must be applied in order to generate gauge-invariant Green
functions, which in turn can be obtained from a gauge-invariant
quantum action $\Gamma$. In this respect, the
background-field-method (BFM) \cite{BFM} is meant to construct
manifest gauge-invariant quantum actions from which well-behaved
Green functions satisfying simple Ward identities can be derived.
This method, implemented at the level of generating functionals,
relies on the decomposition of the gauge fields into two
parts:\footnote{Indeed, every bosonic field must be decomposed into
a quantum and a classic part \cite{BFMSM}.} the quantum field
$A^a_\mu$ and the background (classical) field $\hat{A}^a_\mu$, i.e.
$A^a_\mu \to A^a_\mu +\hat{A}^a_\mu$. In the generating functional
only the quantum fields are integrated out, whereas the background
fields are treated as sources. This means that only the quantum
fields can circulate inside the loops. This method allows one to
introduce a gauge-fixing procedure for the quantum fields without
spoiling the gauge invariance of the quantum action with respect to
the classical fields. Although it is necessary to define a
gauge-fixing procedure for both the quantum and the classical fields
in order to define $S$-matrix elements, only a gauge-fixing scheme
for the quantum gauge fields  is required to define general
off-shell Green functions. The quantum action is invariant under
ordinary gauge transformations of the classical fields, while the
quantum fields transform as the adjoint representation of the group
in consideration. In other words, the so constructed action
$S[A+\hat{A}]$ is degenerate with respect to the background fields
but nondegenerate with respect to the quantum fields. The Green
functions derived from the quantum action $\Gamma[\hat{A}]$ are
gauge invariant in the sense that they satisfy simple Ward
identities, but it is worth stressing that they are still dependent
on the gauge parameter $\xi_Q$ that characterizes the gauge-fixing
scheme used for the quantum fields, and so there is no gauge
independence. The BFM has proved useful in many applications
\cite{BFMAPLIC}, simplifying both technically and conceptually the
calculation of radiative corrections.

As already mentioned, Green functions arising from a conventional
quantum action (a BRST-invariant but gauge-noninvariant one) contain
a lot of unphysical information that is removed once they are
inserted into some physical observable. Some of this unwanted
information can be removed at the level of the generating functional
through the BFM formalism, which allows one to construct a
gauge-invariant quantum action from which gauge-invariant Green
functions can be obtained. Although the resultant Green functions
satisfy simple Ward identities, they are not gauge independent. Thus
far there is still no known mechanism yielding both gauge-invariant
and gauge-independent Green functions directly from the generating
functional, although there is already a diagrammatic method meant
for this purpose, the so-called pinch technique (PT) \cite{PT}. This
method relies on constructing well-behaved Green functions by
combining some individual contributions from self-energies, vertex,
and box diagrams, which usually appear in physical processes. In
general, the Feynman rules used in this diagrammatic approach are
derived from a conventional effective action, though those derived
from the BFM have also been used for a deeper study of the method
self-consistence \cite{PTBFM}. Although the PT was first introduced
for the study of pure Yang-Mills theories at the one-loop level
\cite{PT,PTBFM,4POINTYM}, it has already been applied to theories
with spontaneous symmetry breaking (SSB) \cite{SU(2)}, including the
study of self-energies \cite{2POINT} and trilinear vertices
\cite{3POINT} involving the electroweak gauge bosons. A complete
calculation of the one-loop contribution to the $WWV$ vertex from
the electroweak bosons was presented in Refs. \cite{PTSM1,PTSM2}: it
was intended to show that the vertex functions satisfy a simple Ward
identity, which establishes a relationship between this vertex and
the $W$ self-energy. More specifically, Ref. \cite{PTSM1} discusses
the gauge independence of the form factors associated with the
$WW\gamma$ vertex for off-shell photon and on-shell $W$ bosons.
Afterwards, an important connection between the PT and the BFM was
established \cite{COINCIDENCE} at the one-loop level by showing that
the Green functions calculated via the BFM Feynman rules coincide
with those obtained through the PT for the specific value $\xi_Q=1$.
More recently, the PT was extended to the two-loop level in the
context of both the Yang-Mills \cite{TWOLPTYM} and the electroweak
sectors \cite{TWOLPTEW}, and the one-loop connection to the BFM was
established too. A step toward a nondiagrammatic formulation of the
PT via the powerful Batalin-Vilkovisky quantization method \cite{BV}
was presented in Ref. \cite{PTBV}. This framework was used to
generalize the PT at any order of perturbation theory
\cite{ALLORDERSPT1}, and it was meant to show that the link between
the PT Green functions and those obtained via the BFM along with the
Feynman-t'Hooft gauge remains at all orders of perturbation theory
\cite{ALLORDERSPT1,ALLORDERSPT2}. The reason for such a link remains
a puzzle, though it is worth noting that the Feynman-t'Hooft gauge
yields no unphysical thresholds. Establishing such a  connection at
any order of perturbation theory is very important  for practical
purposes because one can simply use the BFM Feynman-t'Hooft gauge
(BFMFG) to calculate gauge-independent off-shell amplitudes, which
happens to be much less cumbersome than the use of the PT.

Although in  conventional quantization schemes the quantum action of
the theory is not gauge invariant, it is still possible to introduce
gauge invariance with respect to a subgroup of such a theory. This
scheme is particulary useful when the quantum fluctuations of the
gauge fields associated with this subgroup are deemed negligible.
For instance, it would be interesting to assess the virtual effects
of the heavy physics lying beyond the Fermi scale on the SM Green
functions in a $SU_L(2)\times U_Y(1)$-covariant manner, in which
case it is only necessary to introduce a quantization scheme for the
heavy fields since the SM fields would only appear as external legs.
This is indeed the philosophy behind the effective Lagrangian
approach widely used in the context of the electroweak theory, where
it is assumed that the new physics effects must respect the
$SU_L(2)\times U_Y(1)$ symmetry. In a specific theory beyond the SM,
a $SU_L(2)\times U_Y(1)$-invariant effective Lagrangian can be
constructed by introducing a $SU_L(2)\times U_Y(1)$-covariant
gauge-fixing procedure  for the heavy gauge bosons in order to
integrate them out in the generating functional. In analogy with the
BFM, the gauge-fixing procedure for the heavy gauge fields must
involve the $SU_L(2)\times U_Y(1)$-covariant derivative given in the
representation in which the heavy fields transform under this group.
This is the reason why such gauges, which were first introduced by
Fujikawa in the context of the SM \cite{Fujikawa}, are called
nonlinear or covariant gauges. In this case, the $W$ propagator is
defined in a covariant way under the electromagnetic $U_e(1)$ group,
so the vertex functions associated with the $WW\gamma$ interaction
and the $W$ self-energy satisfy a simple Ward identity. The most
general renormalizable structure of this gauge-fixing procedure  has
been discussed from the BRST-symmetry standpoint in \cite{NLGCDTGT},
and a discussion about the difficulties on implementing the
Faddeev-Popov method (FPM) has been presented too \cite{FPM}. This
gauge-fixing procedure has proved a valuable tool in radiative
corrections as it simplifies considerably the loop calculations
\cite{NLGSM}.  The method has also been used to quantize Yang-Mills
theories without SSB \cite{NLGYM}. We will show below that, within
some specific models, it is possible to use a nonlinear gauge to
parametrize in an $SU_L(2)\times U_Y(1)$-invariant way the impact of
new physics on the SM Green functions. In particular, we will show
that it is possible to use this class of gauges to estimate the
one-loop effects of new heavy gauge bosons on the $WWV$ vertex.

We are interested in the sensitivity of the $WWV$ vertex to the new
heavy gauge bosons predicted by the so-called minimal $331$ model
\cite{P,F}, which is based on the $SU_C(3)\times SU_L(3)\times
U_X(1)$ gauge group. Apart from predicting signals of new physics at
the TeV scale, this model introduces unique features that have been
the focus of great interest recently \cite{FEATURES} such as a
possible approach to the solution of the family replication problem.
In this model, the lepton spectrum is the same as the SM one, but it
is accommodated in $SU_L(3)$ antitriplets; the quark sector is also
arranged in the fundamental representation of this group, which
requires the introduction of three new quarks. In order to endow all
the particles with mass, a Higgs sector composed by three triplets
and one sextet of $SU_L(3)$ is required, though only one of the
triplets is needed to break down $SU_L(3)\times U_X)(1)$ into
$SU_L(2)\times U_Y(1)$ at the new physics scale $u>v$. In the first
stage of SSB, there emerge singly and doubly charged gauge bosons in
a doublet of the $SU_L(2)$ group, as well as a new neutral boson
$Z'$. The new charged gauge bosons were dubbed bileptons because
they carry two units of lepton number. The three exotic quarks and a
CP even Higgs boson  do not couple to the $W$ gauge boson since they
emerge as singlets of $SU_L(2)$, and get their mass at the $u$
scale. Thus, at this scale, the $WW\gamma$ and $WWZ$ vertices can
only receive contributions from the bileptons. The fact that the
$SU_L(2)$ group is totally embedded in $SU_L(3)$ gives rise to
unusual couplings between the bileptons and the SM gauge fields,
which arise via the electroweak covariant derivative since the
bileptons transform as the fundamental representation of $SU_L(2)$.
It turns out that these couplings do not involve any mixing angle
and are similar both in strength and Lorentz structure to those
couplings existing between the SM gauge bosons themselves, as
opposed to the gauge bosons appearing in other SM extensions. Our
main goal is to estimate, in a $SU_L(2)\times U_Y(1)$-invariant way,
the sensitivity of the $WWV$ vertex to the bileptons. To this end,
we introduce a $SU_L(2)\times U_Y(1)$-covariant gauge-fixing
procedure for the bileptons, which leads to  an invariant quantum
action. We will show below that the resulting  $WW\gamma$ and $WWZ$
Green functions are gauge invariant and satisfy simple Ward
identities. Another feature worthwhile to emphasize is that the FPM
fails when it is attempted to be used in conjunction with this class
of gauges: the resultant theory is not renormalizable
\cite{NLGCDTGT}. Instead of using this method, we will present a
discussion based on BRST symmetry \cite{BRST}, which is a powerful
formalism adequate not only to quantize Yang-Mills theories with
broader gauge-fixing procedures, as the nonlinear ones, but also to
quantize more general gauge systems. As we will see below, our
quantization scheme incorporates the main ingredient of the BFM,
namely, the gauge invariance of the quantum action, which turns loop
calculations into a somewhat simple task.

The rest of the paper has been organized as follows. In Sec.
\ref{model} a brief description of the minimal $331$ model is
presented. Particular emphasis is given to the Yang-Mills sector. In
Sec. \ref{gauge} a  $SU_L(2)\times U_Y(1)$-covariant gauge-fixing
procedure for the bileptons is presented along with a discussion on
the advantages of using the BRST formalism instead of the FPM. Sec.
\ref{cal} is devoted to present the one-loop amplitudes for the
$WW\gamma$ and $WWZ$ vertices, whereas in Secs. \ref{dis} and
\ref{con} we discuss our results and present the conclusions.

\section{The minimal $331$ model}
\label{model} The $SU_C(3)\times SU_L(3)\times U_X(1)$ model has
been discussed to some extent in the literature
\cite{WWg331,FEATURES,ZPRIME}. We will only focus on those features
that are relevant for the present discussion. In particular, we will
concentrate on the first stage of SSB, when the $331$ group is
broken down into the SM group. The complete Higgs sector is
comprised by three triplets and one sextet of $SU_L(3)$, but only
the following triplet is necessary to break $SU_L(3)\times U_X(1)$
into $SU_L(2)\times U_Y(1)$:

\begin{equation}
\Phi=\left(\begin{array}{ccc} \Phi_Y \\
\phi^0 \end{array}\right): \ \ \ (1,3,1),
\end{equation}
where $\Phi_Y$ is a doublet of $SU_L(2)\times U_Y(1)$ with
hypercharge 3.

In the fundamental representation of $SU_L(3)\times U_X(1)$, the
covariant derivative can be written as

\begin{equation}
{\cal D}_\mu =\partial_\mu
-ig\frac{\lambda^a}{2}A^a_\mu-ig_XX\frac{\lambda^9}{2}X_\mu, \ \ \
(a=1,\cdots,8),
\end{equation}
where $\lambda^a$ $(a=1,\cdots,8)$ are the Gell-Mann matrices and
$\lambda^9=\sqrt{2/3}\,{\rm diag}(1,1,1)$. The generators are
normalized as ${\rm Tr}\lambda^a\lambda^b=2\delta^{ab}$. When $\Phi$
develops a vacuum expectation value,
$\Phi^\dag_{0}=(0,0,u/\sqrt{2})$, the exotic quarks, one physical
neutral scalar, and the new gauge bosons acquire masses, whereas the
remaining scalar multiplets give mass to the SM particles. The first
stage of SSB is accomplished by $\Phi_{0}$ according to the
following scheme: six generators are broken ($\lambda^b\Phi_{0}\neq
0$  for $b=4,\cdots,9$) and the remaining ones leave invariant the
vacuum ($\lambda^a \Phi_{0}=0$ for $a=1,2,3$). Notice that
$\sqrt{3}(\lambda^8+\sqrt{2}X\lambda^9)\Phi_{0}=0$, so the
hypercharge can be identified with the following linear combination
of broken generators:  $Y=\sqrt{3}(\lambda^8+\sqrt{2}X\lambda^9)$.
At this stage of SSB there appear one single charged bilepton and
one doubly charged one defined by

\begin{eqnarray}
Y^{++}_\mu&=&\frac{1}{\sqrt{2}}(A^4_\mu-iA^5_\mu), \\
Y^+_\mu&=&\frac{1}{\sqrt{2}}(A^6_\mu-iA^7_\mu),
\end{eqnarray}
which have the following mass
\begin{equation}
m_{Y^{++}}=m_{Y^+}=m_Y=\frac{gu}{2}.
\end{equation}
According to the quantum number assignment, these fields accommodate
in one doublet of $SU_L(2)\times U_Y(1)$ with hypercharge 3:

\begin{equation}
Y_\mu=\left(\begin{array}{ccc} Y^{++}_\mu \\
Y^+_\mu \end{array}\right).
\end{equation}
As far as the scalar triplet is concerned, the two components of the
$SU_L(2)\times U_Y(1)$ doublet with hypercharge 3, $\Phi_Y$, coincide
with the pseudo-Goldstone bosons associated with the bilepton
doublet:
\begin{equation}
\Phi_Y=\left(\begin{array}{ccc} G^{++}_Y \\
G^+_Y \end{array}\right).
\end{equation}
Finally, the third component of the triplet contains the physical
Higgs boson and the pseudo-Goldstone boson associated with $Z'$:
\begin{equation}
\phi^0=\frac{1}{\sqrt{2}}(u+H^{'0}+iG_{Z'}).
\end{equation}

The bilepton masses receive new contributions at the Fermi scale,
when $SU_L(2)\times U_Y(1)$ is broken into $U_e(1)$, which yields an
upper bound on the splitting between the square bilepton masses:
\begin{equation}
|m^2_{Y^{++}}-m^2_{Y^+}|\leq m^2_W.
\end{equation}
Therefore, $m_{Y^{++}}$ and $m_{Y^+}$ cannot not be very different:
one of them cannot become arbitrarily large while the other one
remains fixed. In fact, the bilepton masses become nearly degenerate
when they are much larger than $m_W$. In addition, the theoretical
constraint $4s^2_W\leq 1$ obtained from matching the gauge couplings
constants at the $SU_L(3)\times U_X(1)$ breaking scale yields an
upper bound on the bilepton masses of the order of 1 TeV
\cite{F,NG}. Therefore, our estimate for the Green functions
associated with the $WWV$ vertex at the $u$ scale would not become
spoiled by the new contributions arising at the $v$ scale. As far as
the remaining gauge bosons are concerned, the gauge fields $A^8_\mu$
and $X_\mu$ mix to produce a massive field $Z'_\mu$, and a massless
gauge boson $B_\mu$\cite{ZPRIME}. The latter is associated with the
$U_Y(1)$ group. At the $u$ scale, the $Z'$ field does not couple
with the $W$ boson, though it can couple to a $W$ boson pair at the
Fermi scale via  $Z'-Z$ mixing \cite{ZPRIME}.

Therefore, although  five massive gauge bosons emerge at the $u$
scale, along with three exotic quarks and one Higgs boson, only the
bileptons couple to the $W$ gauge boson. These interactions are
dictated by the $SU_L(2)\times U_Y(1)$ symmetry, and emerge entirely
from the Yang-Mills sector associated with the $SU_L(3)\times
U_X(1)$ group. The corresponding Lagrangian is composed of the
following three $SU_L(2)\times U_Y(1)$-invariant pieces
\cite{WWg331}:

\begin{equation}
{\cal L}_{YM}=-\frac{1}{4}F^a_{\mu \nu}F^{\mu
\nu}_a-\frac{1}{4}X_{\mu \nu}X^{\mu \nu}={\cal L}_{SM}+{\cal
L}_{SMNP}+{\cal L}_{NP},
\end{equation}
where ${\cal L}_{SM}$ is the electroweak Yang-Mills Lagrangian:

\begin{equation}
{\cal L}_{SM}=-\frac{1}{4}W^i_{\mu \nu}W^{\mu
\nu}_i-\frac{1}{4}B_{\mu \nu}B^{\mu \nu},
\end{equation}
where we have made the association $A^a_\mu \to W^i_\mu$, for
$a=1,2,3$. ${\cal L}_{SMNP}$ encompasses the interactions between
the SM gauge bosons and the new ones, it can be written as

\begin{eqnarray}
\label{LSMNP} {\cal L}_{SMNP}&=&-\frac{1}{2}(D_\mu Y_\nu-D_\nu
Y_\mu)^\dag (D^\mu Y^\nu-D^\nu Y^\mu)-Y^{\dag \mu}(ig{\bf W}_{\mu
\nu}+ig'{\bf
B}_{\mu \nu})Y^\nu \nonumber \\
&&-\frac{ig\sqrt{3}\sqrt{1-4s^2_W}}{2c_W}Z'_\mu[Y^\dag_\nu(D^\mu
Y^\nu-D^\nu Y^\mu)-(D^\mu Y^\nu-D^\nu Y^\mu)^\dag Y_\nu],
\end{eqnarray}
where we have introduced the definitions ${\bf W}_{\mu
\nu}=\tau^iW^i_\mu/2$ and ${\bf B}_{\mu \nu}=YB_{\mu \nu}/2$. In
addition, $D_\mu=\partial_\mu -ig{\bf W}_\mu-ig'{\bf B}_\mu$ stands
for the covariant derivative associated with the electroweak group.
The first two terms of this Lagrangian induce a diversity of
couplings between the SM gauge bosons and the bileptons. Finally,
the term ${\cal L}_{NP}$ induces the interactions between the $Z'$
boson and the bileptons:

\begin{eqnarray}
{\cal L}_{NP}&=&-\frac{1}{4}Z'_{\mu \nu}Z'^{\mu
\nu}-\frac{ig\sqrt{3}\sqrt{1-4s^2_W}}{2c_W}Z'_{\mu \nu}Y^{\dag
\mu}Y^\nu-\frac{3g^2(1-4s^2_W)}{4c^2_W}Z'_\mu Y^\dag_\nu(Z'^\mu
Y^\nu-Z'^\nu Y^\mu)\nonumber \\
&&+\frac{g^2}{2}\Big(Y^\dag_\mu
\frac{\tau^i}{2}Y_\nu\Big)\Big(Y^{\dag
\mu}\frac{\tau^i}{2}Y^\nu-Y^{\dag
\nu}\frac{\tau^i}{2}Y^\mu\Big)+\frac{3g^2}{4}(Y^\dag_\mu
Y_\nu)(Y^{\dag \mu}Y^\nu-Y^{\dag \nu}Y^\mu).
\end{eqnarray}
Note that each term in the last two Lagrangians is separately
invariant under the electroweak group.

As evident from above, the one-loop level contributions to the
$WW\gamma$ and $WWZ$ vertices arise only from the first two terms
appearing in ${\cal L}_{SMNP}$. A gauge-fixing procedure covariant
under the $SU_L(2)\times U_Y(1)$ group will be introduced below for
the bilepton sector.

\section{Gauge-fixing procedure and Feynman rules}
\label{gauge}It has already been mentioned that the FPM fails to
quantize Yang-Mills theories possessing more general supplementary
conditions than the linear ones. This stems from the fact that the
FPM leads to an action which is bilinear in the ghost and antighost
fields since they arise essentially from the integral representation
of a determinant. This is not however the most general situation
that can arise since an action including quartic ghost interactions
at the tree level is still consistent with BRST symmetry and the
power counting criterion of renormalization theory. It turns out
that the FPM does succeed when applied to linear gauges because
quartic ghost interactions cannot arise from loop effects due to
antighost translation invariance,\footnote{Invariance under the
transformation $\bar{C}^a\to \bar{C}^a+c^a$, with $c^a$ arbitrary
constant parameters.} which stems from the fact that the antighost
fields appear only through their derivatives. However, this symmetry
is lost in the case of nonlinear gauges since the gauge-fixing
functions depend on bilinear terms of gauge fields. These terms are
responsible for the presence of ultraviolet-divergent quartic-ghost
interactions at one-loop level. This means that renormalizability
becomes ruined when the FPM is attempted to be used in the context
of nonlinear gauges. It is thus convenient to discard the FPM and
building up instead the most general action consistent with BRST
symmetry and renormalization theory. BRST symmetry arises naturally
from the field-antifield formalism \cite{BV}, which has proved a
powerful tool in quantizing gauge systems. In the case of Yang-Mills
systems, the gauge-fixed BRST action  has a simple structure
\cite{BV}:
\begin{equation}
S_{BRST}=S_{331}+\delta\Psi,
\end{equation}
where $S_{331}$ is the gauge-invariant classical action, $\delta$ is
the BRST operator, and $\Psi$ is the fermion action. $S_{BRST}$
would be nondegenerate if a gauge-fixing procedure for all the gauge
fields of the model was introduced. Since we are only interested in
the virtual effects of the bileptons, a gauge-fixing procedure for
these fields is only necessary. Furthermore, we use a gauge-fixing
procedure covariant under the $SU_L(2)\times U_Y(1)$ group because
we are interested in preserving such a symmetry. The resultant
$S_{BRST}$ action is nondegenerate with respect to the bilepton
fields,\footnote{If necessary, a gauge-fixing procedure for the new
$Z$ boson can be introduced without affecting the $SU_L(2)\times
U_Y(1)$-invariance of $S_{BRST}$.} but degenerate with respect to
the electroweak fields. As a consequence, a $SU_L(2)\times
U_Y(1)$-invariant quantum action can be constructed out of which
gauge-invariant  Green functions, $<0|W^+_\mu (x_1)W^-_\nu
(x_2)A_\lambda (x_3)|0>$ and $<0|W^+_\mu (x_1)W^-_\nu (x_2)Z_\lambda
(x_3)|0>$, satisfying simple Ward identities, can be derived. More
specifically, we introduce a fermion action defined as follows:
\begin{equation}
\Psi=\int
d^4x[\bar{C}^{\bar{a}}(f^{\bar{a}}+\frac{\xi}{2}B^{\bar{a}}+f^{\bar{a}bc}\bar{C}^bC^c)],
\ \ \   \bar{a}=4,5,6,7; \ \ \ b,c=1,\cdots 8,
\end{equation}
where $f^{\bar{a}}$, $\bar{C}^{\bar{a}}$, and $B^{\bar{a}}$ are the
gauge-fixing functions, the antighost fields, and the auxiliary
scalar fields associated with the $A^{\bar{a}}_\mu$ gauge fields,
respectively. In addition, $C^a$ are the ghost fields associated
with the $A^a_\mu$ fields, $f^{abc}$ are the $SU_L(3)$ structure
constants, and $\xi$ is the gauge parameter. Note that the
$f^{\bar{a}bc}\bar{C}^bC^c$ term cannot arise from the FPM, though
its presence is necessary to obtain renormalizability when the
gauge-fixing functions are nonlinear. Using the usual BRST
transformations, we obtain for the $\Psi$ variation
\begin{equation}
\delta \Psi=\int
d^4x\Big[\frac{\xi}{2}B^{\bar{a}}B^{\bar{a}}+(f^{\bar{a}}+2f^{\bar{a}bc}\bar{C}^bC^c)B^{\bar{a}}
-\bar{C}^{\bar{a}}(\delta
f^{\bar{a}})-\frac{1}{2}f^{\bar{a}bc}f^{cde}\bar{C}^{\bar{a}}\bar{C}^bC^dC^e\Big].
\end{equation}
On the other hand, since the auxiliary fields $B^{\bar{a}}$ appear
quadratically, they can be integrated out in the generating
functional. Since the coefficients of the quadratic terms do not
depend on the fields, their integration is equivalent to applying
the equations of motion to the gauge-fixed BRST action. Once these
steps are done, we obtain an action defined by the following
$SU_L(2)\times U_Y(1)$-invariant Lagrangian
\begin{equation}
{\cal L}_{BRST}={\cal L}_{331}+{\cal L}_{GF}+{\cal L}_{FP},
\end{equation}
where ${\cal L}_{331}$ is the gauge invariant Lagrangian of the
$331$ model, whereas ${\cal L}_{GF}$ and ${\cal L}_{FP}$ arise from
the action $\delta \Psi$. The former is the gauge-fixing term, which
can be written as
\begin{equation}
{\cal L}_{GF}=-\frac{1}{2\xi}f^{\bar{a}}f^{\bar{a}},
\end{equation}
and ${\cal L}_{FP}$ represents the ghost sector:
\begin{equation}
{\cal L}_{FP}=-\bar{C}^{\bar{a}}(\delta
f^{\bar{a}})-\frac{2}{\xi}f^{\bar{a}bc}f^{\bar{a}}\bar{C}^bC^c-\frac{1}{2}f^{\bar{a}bc}
f^{cde}\bar{C}^{\bar{a}}\bar{C}^bC^dC^e.
\end{equation}
While the first term in this Lagrangian does arise when the FPM is
used, the remaining ones are new and must be preserved if a
nonlinear function  $f^{\bar{a}}$ is introduced. If a linear gauge
is used, these terms can be removed after invoking
antighost-translation invariance.

\begin{figure}[!htb]
\centering
\includegraphics[width=2.5in]{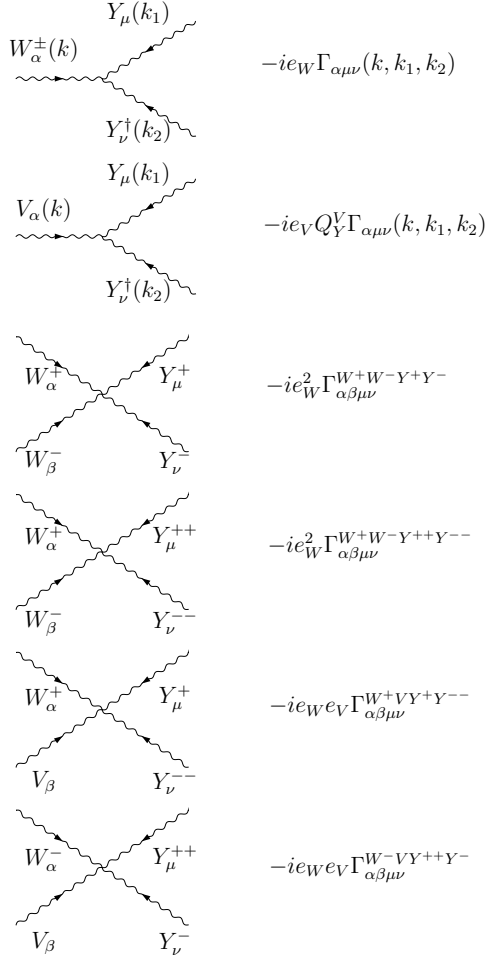}
\caption{\label{FIG1} Feynman rules for the trilinear and quartic
vertices involving the bileptons and SM gauge fields in the
$SU_L(2)\times U_Y(1)$-covariant $R_\xi$-gauge.}
\end{figure}

We are now ready to introduce the most general $SU_L(2)\times
U_Y(1)$-covariant $f^{\bar{a}}$ functions, and we will take
advantage of the fact that every coupling involving at least one
pseudo-Goldstone boson and every coupling with gauge freedom can be
modified leaving unaltered the $S$ matrix. The most general
renormalizable gauge-fixing functions consistent with this symmetry
can be written as
\begin{equation}
f^{\bar{a}}=(\delta^{\bar{a}b}\partial_\mu-gf^{\bar{a}bi}A^i_\mu)A^{\mu
b}-\frac{\xi g}{\sqrt{3}}f^{\bar{a}b8}\Phi^\dag \lambda^b \Phi, \
\ \ \bar{a}=4,5,6,7; \ \ \ i=1,2,3,8.
\end{equation}
Notice that these gauge-fixing functions are nonlinear in both the
vector and the scalar sectors. To be fully aware of the covariant
structure of these gauge-fixing functions, it is convenient to
express them in terms of the mass eigenstates fields. Using the
definitions
\begin{eqnarray}
f^{++}_Y&=&\frac{1}{\sqrt{2}}(f^4-if^5), \\
f^{+}_Y&=&\frac{1}{\sqrt{2}}(f^6-if^7),
\end{eqnarray}
we can write
\begin{equation}
f_{Y}=\left(\begin{array}{ccc} f^{++}_Y \\
f^+_Y
\end{array}\right)=\Big(D_\mu-\frac{ig\sqrt{3}\sqrt{1-4s^2_W}}{2c_W}Z'_\mu\Big)Y^\mu-
\frac{ig\xi}{\sqrt{2}}\phi^{0*}\Phi_Y,
\end{equation}
where $D_\mu$ is the $SU_L(2)\times U_Y(1)$-covariant derivative
given in the doublet representation. From this expression, it is
evident that $f_Y$ transforms as $Y_\mu$ or $\Phi_Y$, i.e. as a
doublet of $SU_L(2)\times U_Y(1)$ with hypercharge 3. As a
consequence, the gauge-fixing term ${\cal L}_{GF}$ is manifestly
invariant under this group. As will become evident below, it is
convenient to decompose the Lagrangian ${\cal L}_{GF}$ into three
gauge-invariant terms:
\begin{equation}
{\cal L}_{GF}={\cal L}_{GF1}+{\cal L}_{GF2}+{\cal L}_{GF3},
\end{equation}
where
\begin{eqnarray}
{\cal L}_{GF1}&=&-\frac{1}{\xi}(D_\mu Y^\mu)^\dag (D_\nu
Y^\nu)-\frac{\xi g^2}{2}(\phi^{0*}\phi^0)(\Phi^\dag_Y\Phi_Y), \\
{\cal L}_{GF2}&=&\frac{ig}{\sqrt{2}}\Big[\phi^{o*}(D_\mu
Y^\mu)^\dag \Phi_Y-\phi^0\Phi^\dag_Y(D_\mu Y^\mu)\Big],\\
{\cal L}_{GF3}&=&\frac{ig\sqrt{3}\sqrt{1-4s^2_W}}{2c_W\xi}Z'_\mu
[(D_\nu Y^\nu)^\dag Y^\mu-Y^{\mu \dag}(D_\nu Y^\nu)]\nonumber \\
&&-\frac{g^2\sqrt{3}\sqrt{1-4s^2_W}}{2\sqrt{2}c_W}Z'_\mu(\phi^{0*}Y^{\mu
\dag}\Phi_Y+\phi^0 \Phi^\dag_Y Y^\mu)\nonumber \\
&&-\frac{3g^2(1-4s^2_W)}{4c^2_W\xi}Z'_\mu Z'_\nu Y^{\mu
\dag}Y^\nu.
\end{eqnarray}
We would like to discuss the dynamics of ${\cal L}_{GF1}$ and ${\cal
L}_{GF2}$. The first term appearing in ${\cal L}_{GF1}$, which is
invariant under the electroweak group, not only allows to define the
bilepton propagators but also modifies nontrivially the couplings
between the bileptons and the electroweak gauge bosons appearing in
the ${\cal L}_{NPSM}$ Lagrangian. When the two Lagrangians are
combined, they lead to trilinear and quartic vertices:
\begin{eqnarray}
{\cal L}_{WYY}&=&ie_W\Big\{W^{+\mu}(Y^{--}_{\mu
\nu}Y^{+\nu}-Y^+_{\mu \nu}Y^{--\nu})-W^+_{\mu
\nu}Y^{--\mu}Y^{+\nu} -W^{-\mu}(Y^{++}_{\mu \nu}Y^{-\nu}-Y^-_{\mu
\nu}Y^{++\nu})+W^-_{\mu \nu}Y^{++\mu}Y^{-\nu}\nonumber \\
&&+\frac{1}{\xi}\Big[W^{+\mu}(Y^+_\mu \partial_\nu
Y^{--\nu}-Y^{--}_\mu \partial_\nu Y^{+\nu})-W^{-\mu}(Y^-_\mu
\partial_\nu Y^{++\nu}-Y^{++}_\mu \partial_\nu Y^{-\nu})\Big]\Big
\},
\end{eqnarray}

\begin{eqnarray}
{\cal L}_{VYY}&=&ie_V\Big \{Q^V_{Y^+}\Big [V^\mu (Y^-_{\mu
\nu}Y^{+\nu}-Y^+_{\mu \nu}Y^{-\nu})-V_{\mu
\nu}Y^{-\mu}Y^{+\nu}+\frac{1}{\xi}V^{\mu}(Y^+_\mu \partial_\nu
Y^{-\nu}-Y^-_\mu \partial_\nu Y^{+\nu})\Big]+\nonumber \\
&&Q^V_{Y^{++}}\Big [V^\mu (Y^{--}_{\mu \nu}Y^{++\nu}-Y^{++}_{\mu
\nu}Y^{--\nu})-V_{\mu
\nu}Y^{--\mu}Y^{++\nu}+\frac{1}{\xi}V^{\mu}(Y^{++}_\mu
\partial_\nu Y^{--\nu}-Y^{--}_\mu \partial_\nu Y^{++\nu})\Big],
\end{eqnarray}

\begin{eqnarray}
{\cal L}_{WWYY}&=&-e^2_W\Big \{W^-_\mu W^{+\mu}(Y^-_\nu
Y^{+\nu}+Y^{--}_\nu Y^{++\nu})\nonumber \\
&&-W^-_\mu
W^+_\nu(2Y^{+\mu}Y^{-\nu}-Y^{-\mu}Y^{+\nu}+2Y^{--\mu}Y^{++\nu}-Y^{--\nu}Y^{++\mu})\nonumber
\\
&&+\frac{1}{\xi}W^-_\mu W^+_\nu
(Y^{-\mu}Y^{+\nu}+Y^{--\nu}Y^{++\mu})\Big\},
\end{eqnarray}

\begin{eqnarray}
{\cal L}_{VWYY}&=&-e_We_VV^\mu\Big \{(Q^V_{Y^+}+Q^V_{Y^{++}})
\Big[Y^{--\nu}(W^+_\mu Y^+_\nu-W^+_\nu Y^+_\mu)+Y^{++\nu}(W^-_\mu
Y^-_\nu-W^-_\nu Y^-_\mu)\Big]\nonumber \\
&&+(Q^V_{Y^{++}}-2Q^V_{Y^+}) \Big[W^{+\nu}(Y^{--}_\mu
Y^+_\nu-Y^{--}_\nu Y^+_\mu)+W^{-\nu}(Y^{++}_\mu Y^-_\nu-Y^{++}_\nu
Y^-_\mu)\Big]\nonumber \\
&&+\frac{1}{\xi} \Big[Q^V_{Y^+}(Y^{--} _\nu W^{+\nu}Y^+_\mu
+Y^{++}_\nu W^{-\nu} Y^-_\mu)+Q^V_{Y^{++}}(Y^{--}_\mu W^+_\nu
Y^{+\nu}+Y^{++}_\mu W^-_\nu Y^{-\nu})\Big]\Big \},
\end{eqnarray}
where $e_V=e$, $Q^V_{Y^+}=1$, and $Q^V_{Y^{++}}=2$ for $V=\gamma$,
whereas $e_V=g/(2c_W)$, $Q^V_{Y^+}=-(1+2s^2_W)$, and
$Q^V_{Y^{++}}=1-4s^2_W$, for $V=Z$. Also, $e_W=g/\sqrt{2}$ in any
case. The respective vertex functions are shown in Fig. \ref{FIG1}.
It is worth emphasizing that, as required by $SU_L(2)\times U_Y(1)$
symmetry, the vertex functions associated with the trilinear
couplings $WYY$ and $VYY$ share the same Lorentz structure:
\begin{equation}
\Gamma_{\alpha \mu \nu}(k,k_1,k_2)=(k_2-k_1)_\alpha g_{\mu
\nu}+(k-k_2-\frac{1}{\xi} k_1)_\mu g_{\alpha
\nu}-(k-k_1-\frac{1}{\xi} k_2)_\nu g_{\alpha \mu}.
\end{equation}
It is not hard to show that it satisfies the following simple Ward
identity
\begin{equation}
k^\alpha \Gamma_{\alpha \mu \nu}(k,k_1,k_2)=\Pi^{Y^\dag
Y^\dag}_{\mu \nu}(k_2)-\Pi^{YY}_{\mu \nu}(k_1),
\end{equation}
where $\Pi^{Y\dag Y\dag}_{\mu \nu}(k_2)$ and $\Pi^{Y Y}_{\mu
\nu}(k_1)$ are two--point vertex functions given by
\begin{equation}
\Pi^{YY}_{\mu \nu}(k)=(-k^2+m^2_Y)g_{\mu
\nu}-\Big(\frac{1}{\xi}-1\Big)k_\mu k_\nu.
\end{equation}
As far as the quartic vertices are concerned, they are characterized
by the following vertex functions:
\begin{equation}
\Gamma^{WWYY}_{\alpha \beta \mu \nu}=g_{\alpha \beta}g_{\mu
\nu}-2g_{\alpha \nu}g_{\beta
\mu}+\Big(1+\frac{1}{\xi}\Big)g_{\alpha \mu}g_{\beta \nu},
\end{equation}

\begin{equation}
\Gamma^{VWYY}_{\alpha \beta \mu
\nu}=(Q^V_{Y^+}+Q^V_{Y^{++}})(g_{\alpha \beta}g_{\mu
\nu}-g_{\alpha \nu}g_{\beta \mu})+3\delta_{VZ}(g_{\alpha
\mu}g_{\beta \nu}-g_{\alpha \nu}g_{\beta
\mu})+\frac{1}{\xi}(Q^V_{Y^+}g_{\alpha \nu}g_{\beta
\mu}+Q^V_{Y^{++}}g_{\alpha \mu}g_{\beta \nu}).
\end{equation}

On the other hand, the scalar part of ${\cal L}_{GF1}$ allows  one
to define unphysical masses for the pseudo-Goldstone bosons
associated with the bileptons, and it also modifies some unphysical
couplings of the Higgs potential.

\begin{figure}
\centering
\includegraphics[width=3in]{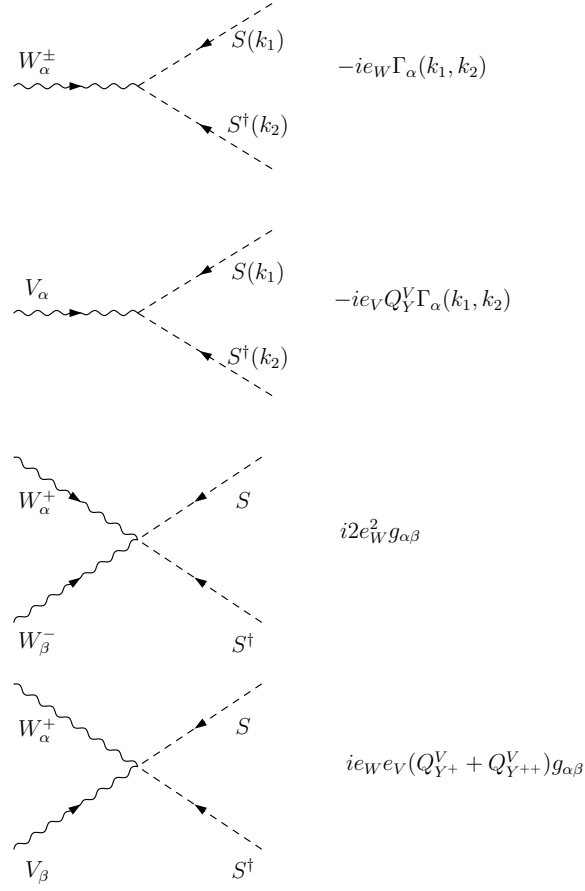}
\caption{\label{FIG2} Feynman rules for the trilinear and quartic
vertices involving SM gauge fields and scalar unphysical particles
(pseudo-Goldstone bosons and ghosts) in the $SU_L(2)\times
U_Y(1)$-covariant $R_\xi$-gauge. In this gauge, the $W$ and $V$
couplings to pseudo-Goldstone bosons and ghosts coincide.}
\end{figure}

It is worthwhile to discuss the dynamics of the ${\cal L}_{GF2}$
Lagrangian. This term has a strong impact on the Higgs
kinetic-energy sector associated with the $\Phi$ triplet as it helps
to remove some unphysical vertices. Note that the latter can be
decomposed into three $SU_L(2)\times U_Y(1)$-invariant pieces:
\begin{equation}
({\cal D}_\mu \Phi)^\dag ({\cal D}^\mu \Phi)={\cal L}_{K\Phi
1}+{\cal L}_{K\Phi 2}+{\cal L}_{K\Phi 3},
\end{equation}
where
\begin{equation}
{\cal L}_{K\Phi 1}=(D_\mu \Phi_Y)^\dag (D^\mu \Phi_Y)+\partial_\mu
\phi^{0*}\partial^\mu
\phi^0+\frac{g^2}{2}[\phi^{0*}\phi^0Y^\dag_\mu Y^\mu
+(\Phi^\dag_YY_\mu)(Y^{\mu \dag}\Phi_Y)],
\end{equation}
and
\begin{equation}
{\cal L}_{K\Phi2}=ie_W\Big[\phi^{0*}Y^\dag_\mu (D^\mu
\Phi_Y)+\Phi^\dag_YY_\mu
\partial^\mu \phi^0-{\rm H.c.}\Big].
\end{equation}
The ${\cal L}_{K\Phi3}$ Lagrangian is not relevant for the present
discussion as it is only composed by the interactions involving the
$Z'$ boson, so we refrain from presenting it here. The Lagrangian
${\cal L}_{K\Phi1}$ gives rise to the interactions between the
pseudo-Goldstone bosons and the electroweak gauge bosons. These
interactions, which are dictated by the electroweak group,
contribute to the $WW\gamma$ and $WWZ$ couplings at the one-loop
level, so the corresponding Feynman rules are shown in Fig.
\ref{FIG2}. On the other hand, the ${\cal L}_{K\Phi2}$ term is
responsible for the appearance of the bilinear terms $Y^{\pm
\pm}_\mu G^{\mp \mp}_Y$ and $Y^{\pm }_\mu G^{\mp }_Y$, as well as
the unphysical trilinear and quartic couplings $Y^{\pm \pm}_\mu
W^\mp G^\mp_Y$, $Y^\pm_\mu W^\pm G^{\mp \mp}_Y$, $H'Y^{\pm \pm}_\mu
W^\mp G^\mp_Y$, and $H'Y^\pm_\mu W^\pm G^{\mp \mp}_Y$. When the
${\cal L}_{GF2}$ Lagrangian is taken into account, all these
couplings vanish. In fact, after adding up these two terms, we
obtain
\begin{equation}
 {\cal L}_{K\Phi2} +{\cal
L}_{GF2}=ie_W\Big[\phi^{0*}\partial_\mu (Y^{\mu
\dag}\Phi_Y)+\Phi^\dag_YY_\mu \partial^\mu\phi^0-{\rm H.c.}\Big],
\end{equation}
where some surface terms were ignored. Needless to say that the
absence of these unphysical vertices renders great simplicity for
some loop calculations.

As far as the ghost sector is concerned, the following definitions
for the ghost fields
\begin{eqnarray}
C^{\pm \pm }_Y&=&\frac{1}{\sqrt{2}}(C^4\mp iC^5), \\
C^\pm_Y&=&\frac{1}{\sqrt{2}}(C^6\mp iC^7),
\end{eqnarray}
and similar expressions for the antighost fields, allow us to
express the corresponding Lagrangian as follows:
\begin{eqnarray}
{\cal L}_{FP}&=&(D_\mu C_Y)^\dag (D^\mu
\bar{C}_Y)+\frac{g^2}{4}\Big[(Y^\dag _\mu \sigma^i
Y^\mu)(C^\dag_Y\sigma^i \bar{C}_Y)+3(Y^\dag_\mu
Y^\mu)(C^\dag_Y\bar{C}_Y)-4(Y^\dag_\mu C_Y)(Y^{\mu
\dag}\bar{C}_Y)\Big]\nonumber \\
&&+\frac{ig}{\sqrt{2}}Y^\dag_\mu M_CD^\mu
\bar{C}_Y+\frac{ig}{2}Y^\dag_\mu {\cal M}_C\bar{C}_Y-\frac{\xi
g}{2}\Big[\phi^{0*}\phi^0C^\dag_Y\bar{C}_Y+\phi^0\Phi^\dag_YM_C\bar{C}_Y-(C^\dag_Y\Phi_Y)(\Phi^\dag_Y\bar{C}_Y)\Big]\nonumber
\\
&&+\frac{i\sqrt{2}}{\xi}\Big[(\bar{M}_CC_Y+M_C\bar{C}_Y)^\dag
(D_\mu Y^\mu)-(D_\mu
Y^\mu)(\bar{M}_CC_Y+M_C\bar{C}_Y)\Big]\nonumber \\
&&-g\Big[\Phi^\dag_Y(\bar{M}_CC_Y+M_C\bar{C}_Y)\phi^0+\phi^{0*}(\bar{M}_CC_Y+M_C\bar{C}_Y)^\dag
\Phi_Y\Big]+{\rm H.c.}\nonumber \\
&&-\frac{1}{2}f^{\bar{a}bc}f^{cde}\bar{C}^{\bar{a}}\bar{C}^bC^dC^e,
\end{eqnarray}
where
\begin{equation}
C_Y=\left(\begin{array}{ccc} C^{++}_Y \\
C^+_Y
\end{array}\right) \ \ \ \ \ \bar{C}_Y=\left(\begin{array}{ccc} \bar{C}^{++}_Y \\
\bar{C}^+_Y
\end{array}\right).
\end{equation}
They have the same quantum numbers that $Y_\mu$ and $\Phi_Y$. In
addition,
\begin{equation}
M_C=\left(\begin{array}{ccc} \frac{1}{\sqrt{2}}(C^3+\sqrt{3}C^8) & \frac{1}{\sqrt{2}}(C^1-iC^2) \\
 \frac{1}{\sqrt{2}}(C^1+iC^2)&-\frac{1}{\sqrt{2}}(C^3-\sqrt{3}C^8)
\end{array}\right),
\end{equation}

\begin{equation}
{\cal M}_C=\left(\begin{array}{ccc} ({\cal D}^{3i}_\mu +\sqrt{3}{\cal D}^{8i}_\mu)C^i &
({\cal D}^{1i}_\mu-i{\cal D}^{21}_\mu)C^i \\
 ({\cal D}^{1i}_\mu +i{\cal D}^{2i}_\mu)C^i&-({\cal
 D}^{3i}_\mu-\sqrt{3}{\cal D}^{8i}_\mu)C^i
\end{array}\right),
\end{equation}
where $i=1,2,3,8$ and ${\cal
D}^{ij}_\mu=\delta^{ij}\partial_\mu-gf^{ija}A^a_\mu $ stands for the
covariant derivative given in the adjoint representation of
$SU_L(3)$. The $\bar{M}_C$ matrix is obtained from $M_C$ after
replacing the ghost fields by antighost fields. Under the
electroweak group, $M_C$ transforms as $M_C \to UM_CU^\dag$, with $U
\in SU_L(2)\times U_Y(1)$. A similar transformation holds for
$\bar{M}_C$ and ${\cal M}_C$. As a consequence, ${\cal L}_{FP}$ is
invariant under the $SU_L(2)\times U_Y(1)$ group.

Both the pseudo-Goldstone boson and the ghost sectors contribute to
the vertex $WWV$  via trilinear and quartic couplings. As shown in
Fig. \ref{FIG2}, the Feynman rules arising from each sector are
identical because each sector is $SU_L(2)\times U_Y(1)$ invariant by
its own. As a consequence, the trilinear vertices $WS^\dag S$ and
$VS^\dag S$ satisfy simple Ward identities:
\begin{equation}
k^\alpha \Gamma^{VS^\dag S}_\alpha=\Pi^{S^\dag
S^\dag}(k_2)-\Pi^{SS}(k_1),
\end{equation}
where $\Gamma^{VS^\dag S}_\alpha=(k_1-k_2)_\alpha$, $S$ stands for a
commutative (pseudo-Goldstone boson) or anticommutative (ghost)
charged scalar, and $\Pi^{SS}(k_i)$ stands for the two-point vertex
functions $\Pi(k_i)=k^2_i-\xi m^2_Y$.

\section{The one-loop $WWV$ vertex}
\label{cal}
\begin{figure}
\centering
\includegraphics[width=2in]{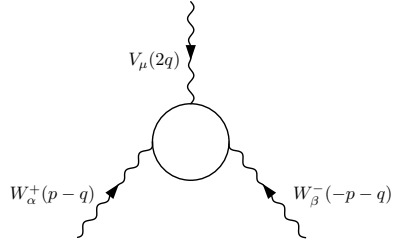}
\caption{\label{FIG3} The trilinear $WWV$ vertex. The large circle
denotes loop contributions and the arrows denote the flow of
momenta.}
\end{figure}

The prospect of the NLC and CLIC \cite{CLIC}, have
triggered the interest in the $e^+e^- \to W^+W^-$ reaction, and
motivated by this we will focus on the $WWV$ vertex with the two
$W$s on-shell and $V$ off-shell. Even in this case there is no
reason to expect a gauge-invariant amplitude arising from a
conventional quantization scheme. We will show below that this is
indeed the case. Retaining just the transverse degrees of freedom of
$V$, the vertex function for the $WWV$ coupling can be written as
\cite{Bardin,Vertex}:
\begin{eqnarray}
\Gamma^V_{\alpha \beta \mu}&=&-ig_V\Big\{A[2p_\mu g_{\alpha
\beta}+4(q_\beta g_{\alpha \mu}-q_\alpha g_{\beta \mu})]+2\Delta
\kappa_V (q_\beta g_{\alpha \mu}-q_\alpha g_{\beta \mu})
+\frac{4\Delta Q_V}{m^2_W}\Big(p_\mu q_\alpha
q_\beta-\frac{1}{2}q^2p_\mu g_{\alpha \beta}\Big)\Big\},
\end{eqnarray}
where
\begin{equation}
g_V=e_V(Q^V_{Y^{++}}-Q^V_{Y^+})=\left\{\begin{array}{ll} gs_W,& V=\gamma \\
 gc_W,& V=Z.
\end{array}\right.
\end{equation}
We have dropped the CP-odd terms since they do not arise at the
one-loop level in the minimal $331$ model. Our notation and
conventions are depicted in Fig. \ref{FIG3}. In the SM, the tree
level values are $A=1$, $\Delta \kappa =0$, and $\Delta Q=0$. In the
case of the on-shell $WW\gamma$ vertex, $\Delta \kappa_\gamma$ and
$\Delta Q_\gamma$ are related to the $W$ magnetic dipole moment
$\mu_W$ and the electric quadrupole moment $Q_W$:
\begin{eqnarray}
\mu_W&=&\frac{e}{2m_W}(2+\Delta \kappa_\gamma), \\
Q_W&=&-\frac{e}{m^2_W}(1+\Delta \kappa_\gamma+\Delta Q_\gamma).
\end{eqnarray}

As already noted, we are interested in the impact of the bileptons
on the $\Delta \kappa_V$ and $\Delta Q_V$ form factors, which
characterize the radiative corrections to the $WWV$ vertex. Our
results will only be an estimate since we are only considering the
new physics effects at the $u$ scale, but our approach has the
advantage of invariance under the $SU_L(2)\times U_Y(1)$ group,
which greatly simplifies the calculations.  At the $u$ scale, only
the bileptons couple to the $W$ boson since they arise as a doublet
of the electroweak group. We already presented the Feynman rules
given in a $R_\xi$-gauge scheme and their $SU_L(2)\times
U_Y(1)$-covariant nature was displayed via simple Ward identities.
The one-loop amplitude of the $WWV$ vertex will also be gauge
invariant, though  gauge dependent, which also occurs when the BFM
is applied. In other words, gauge-invariant quantum actions render
gauge-invariant but not gauge-independent Green functions. However,
motivated by the link between the BFMFG and the PT, we will present
our results in the Feynman-t'Hooft gauge.

\begin{figure}
\centering
\includegraphics[width=3in]{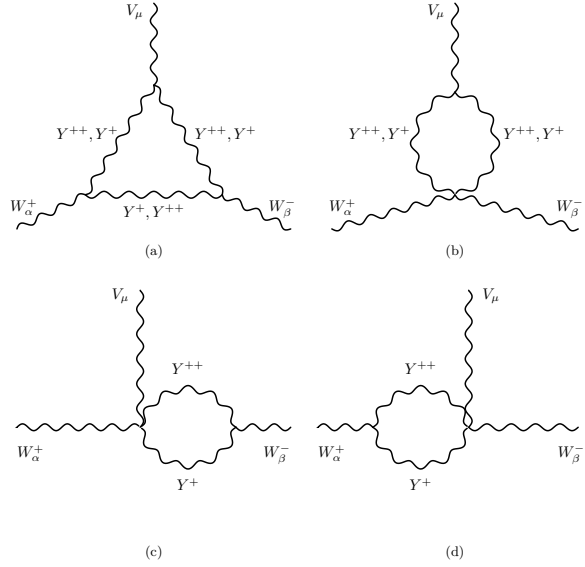}
\caption{\label{FIG4} Feynman diagrams for the $WWV$ vertex in the
$SU_L(2)\times U_Y(1)$-covariant gauge. The pseudo-Goldstone bosons
and ghosts contribute through an identical set of diagrams, but only
the triangle ones give a nonvanishing contribution to the form
factors $\Delta \kappa_V$ and $\Delta Q_V$.}
\end{figure}

The generic Feynman diagrams contributing to the $WWV$ vertex are
shown in Fig. \ref{FIG4}. It turns out that the bileptons contribute
through all these diagrams, but the only nonvanishing contribution
of scalar particles arises from the triangle graphs. In addition,
owing to the separate $SU_L(2)\times U_Y(1)$ invariance of the ghost
and scalar sectors, the ghost-antighost contribution is exactly
minus twice the one coming from the pseudo-Goldstone bosons. Taking
into account all these facts, the total amplitude can be written as
\begin{equation}
\Gamma^V_{\alpha \beta \mu}=-g_VI_{\alpha \beta \mu},
\end{equation}
with $I_{\alpha \beta \mu}$ the loop amplitude, which is the same
for both $WW\gamma$ and $WWZ$ vertices. It is worth mentioning that
the associated Green functions differ only by the factor $g_V$, just
as occurs at the tree-level. This means that $SU_L(2)\times U_Y(1)$
invariance is preserved at the one-loop level. Thus the loop
amplitude $I_{\alpha \beta \mu}$ for on-shell $W$ bosons must
satisfy the simple Ward identity
\begin{equation}
q^\mu I_{\alpha \beta \mu}=0,
\end{equation}
which can be verified once the loop integrals  are solved
explicitly. The explicit calculation shows that the $\Delta
\kappa_V$ and $\Delta Q_V$ form factors are given by
\begin{eqnarray}
\Delta
\kappa_V&=&\frac{6\,a}{(4x_W-1)^3}\Bigg\{x_W(4x_W-1)(8x_W+3)-6x_W\Big[x_W(1+x_W)+3x_Y(1-4x_W)\Big]Q^2C_0 \nonumber \\
&+&4x_Y (4x_W-1)^2\Big[B_0(3)-B_0(1)\Big]+\left[26x^2_W+32x_Y
x_W(1-4x_W)+x_W\right]\Big[B_0(1)-B_0(2)\Big]\Bigg\},
\end{eqnarray}

\begin{eqnarray}
\Delta
Q_V&=&\frac{12\,a}{(4x_W-1)^3}\Bigg\{6x_W\Big[2x_W\Big(2x^3_W-2x^2_W(1+4x_Y)+x_W(1+6x_Y)-3x_Y\Big)+x_Y\Big]Q^2C_0\nonumber
\\
&+&4x_W\left(x_W(6x^2_W-5x_W+8x_Y-1)-2x_Y\right)\Big[B_0(1)-B_0(2)\Big]+4x_Wx_Y(4x_W-1)^2\Big[B_0(2)-B_0(3)\Big]
\nonumber \\
&-&x_W(4x_W-1)\left(1+2x_W(6x_W-1)\right)\Bigg\},
\end{eqnarray}
where we have introduced the definitions $Q=2q$, $a=g^2/96\pi^2$,
$x_W=m^2_W/Q^2$, and $x_Y=m^2_Y/Q^2$. $B_0(i)$ and $C_0$ stand for
the following Passarino-Veltman scalar functions:
$B_0(1)=B_0(m^2_W,m^2_Y,m^2_Y)$, $B_0(2)=B_0(Q^2,m^2_Y,m^2_Y)$,
$B_0(3)=B_0(0,m_Y^2,m_Y^2)$, and
$C_0=C_0(Q^2,m^2_W,m^2_W,m^2_Y,m^2_Y,m^2_Y)$.

\section{Discussion}
\label{dis}The $e^+e^-\to W^+W^-$ reaction will play an essential
role in future researches at $e^+e^-$ colliders: it will provide
relevant information for our knowledge of the SM such as a more
precise determination of the $W$ mass and its width decays, and it
will also open up the possibility for detecting new physics effects
via the distinctive $s$-channel contribution from the $WWV$ vertex.
In the SM, the radiative corrections to the $e^+e^-\to W^+W^-$
process have been widely studied both for on-shell \cite{RCSMON} and
off-shell $W$ gauge bosons \cite{RCSMOFF}. New physics effects have
also been studied in a model-independent manner using the effective
Lagrangian approach \cite{ELS}. Beyond the SM, the reaction
$e^+e^-\to W^+W^-$ has been analyzed in technicolor theories
\cite{TC} and supersymmetric models \cite{SSM}.

As for the radiative corrections to the $WWV$ vertex, they have
received considerably attention. In the SM, the one-loop amplitudes
were calculated using the conventional quantization scheme along
with the Feynman-t'Hoof gauge \cite{TLVSMC}. As emphasized in that
work, the resultant amplitudes are not gauge invariant, which is
evident from the presence of infrared divergences and the bad
high-energy behavior of the $\Delta \kappa_V$ form factor
\cite{TLVSMC}. In contrast, it was found that $\Delta Q$ is
well-behaved. Shortly afterwards, these vertices were revisited by
Papavassiliou and Philippides \cite{TLVSMPT} in a gauge-invariant
way via the PT. The form factor $\Delta \kappa_V$ obtained by these
authors disagrees from that presented in \cite{TLVSMC}, though there
is agreement for $\Delta Q_V$. It was found that the radiative
corrections to $\Delta \kappa_V$ are of the order of $\alpha/\pi$,
whereas $\Delta Q_V$ is about one order of magnitude below. For
instance, $\Delta \kappa_\gamma$ goes from $10^{-3}$ for $Q=200$ GeV
to $10^{-4}$ for $Q=1000$ GeV \cite{TLVSMPT}, whereas $\Delta
Q_\gamma$ ranges from $10^{-4}$ to $6\times 10^{-5}$ in the same
energy range \cite{TLVSMC,TLVSMPT}. As far as experimental
measurements are concerned, the constraint $|\Delta
\kappa_V|,|\Delta Q_V|\lesssim 1/2$ was obtained from CERN
\cite{BCERN} and Fermilab data \cite{BFERMILAB}. It is expected that
this constraint is substantially improved at the CERN large hadron
collider (LHC). Even more, it has been argued that a deviation at
the $10^{-3}$ level might be measured at NLCs. As a consequence,
only $\Delta \kappa_V$ would be at the reach of NLCs, though the
appearance of new physics effects may improve this situation. This
is not the case however for one of the more popular SM extensions,
namely, supersymmetry, which yields similar or smaller contributions
than the SM ones \cite{TLVMSSM}.

We turn now to our results. The $\Delta \kappa_V$ and $\Delta Q_V$
form factors depend on $Q^2$, $m_Y$, and $m_W$. As for $Q^2$, it can
take both positive (time-like) and negative (space-like) values.
However, motivated by the prospect of NLCs, the form factors will be
evaluated for $Q>100$ GeV. It is also evident that for relative low
energies, where the SM  contribution is dominant, those
contributions of very heavy bileptons will be highly suppressed.
However, when $Q$ and $m_Y$ are of the same order, it is expected
that the new physics contributions become more relevant. One
interesting feature of the minimal $331$ model is the constraint
$s^2_W<1/4$ obtained from theoretical arguments, which in turn
translates into $m_Y\lesssim 1.5$ TeV in the minimal 331 model
\cite{NG}, though it can be relaxed by introducing a more complex
Higgs sector. We will consider this upper constraint for the
bilepton mass. On the other hand, the most stringent lower bound
$m_Y>850$ GeV \cite{LBMAM} arises from muonium-antimuonium
conversion. It has been argued however that this bound can be evaded
in a more general context since it relies on very restrictive
assumptions \cite{Pleitez}. Other strong limit, $m_Y>750$ GeV,
arises from fermion pair production and lepton flavor violating
decays \cite{LBLFV}, and the bound $m_{Y^\pm}>440$ GeV, valid only
for the singly charged bilepton mass, was derived from limits on the
muon decay width \cite{LBMD}. We would like to stress that all these
bounds depend on several assumptions, and in principle there still
remains the possibility of lighter bileptons. Although it is quite
unlikely the existence of a relatively light bilepton, as way of
illustration, we will concentrate on the range $2m_W<m_Y<12m_W$.
Also, although the contribution of a  bilepton is expected to become
more significant when its mass is of the same order of magnitude
than that of $Q$, we will present results in the range 100 GeV
$<Q<1000$ GeV.

\begin{figure}[!htb]
\centering
\includegraphics[width=4in]{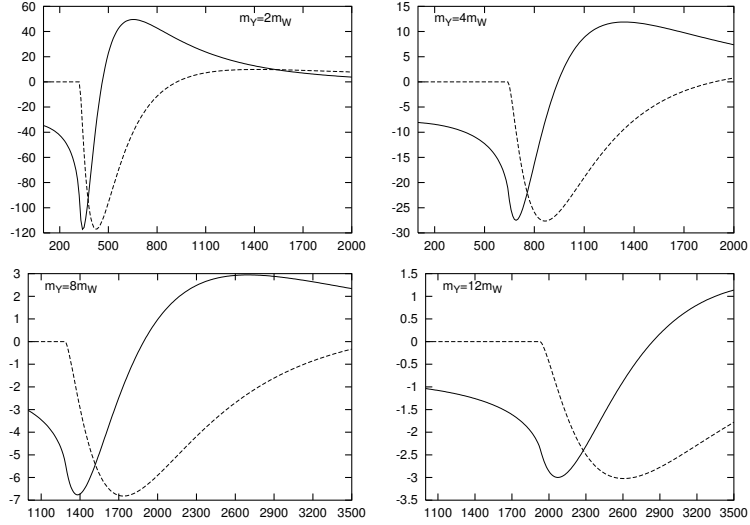}
\caption{\label{DeltaK} The $\Delta \kappa_V$ form factor, in units
of $10^{-6}$, as a function of the center-of-mass energy
$|Q|=\sqrt{s}$, in units of GeV, for several values of $m_Y$. The
solid line is for the real part of $\Delta \kappa_V$ and the dashed
line for its imaginary part.}
\end{figure}

\begin{figure}[!htb]
\centering
\includegraphics[width=4in]{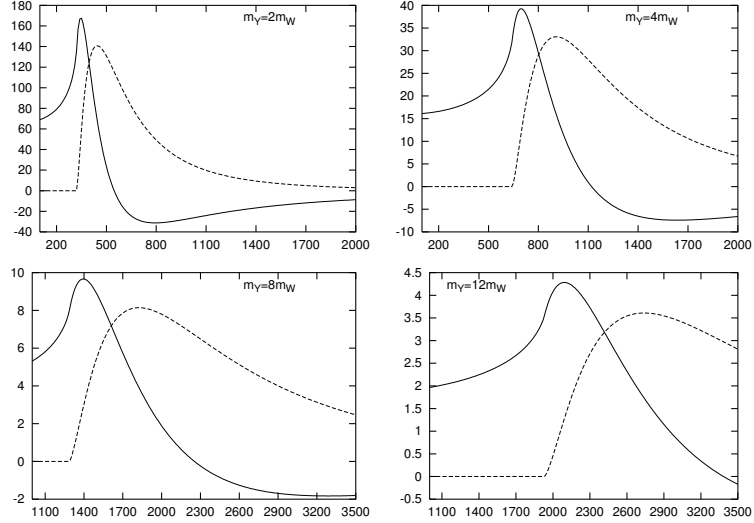}
\caption{\label{DeltaQ} The same as in Fig. \ref{DeltaK} but for the
$\Delta Q_V$ form factor.}
\end{figure}

The $\Delta \kappa_V$ and $\Delta Q$ form factors are shown in Figs.
\ref{DeltaK} and \ref{DeltaQ} for $m_Y=2m_W$, $4m_W$, $8m_W$, and
$12m_W$. We have included the values that arise above the threshold
$Q\geq 2m_Y$ only by completeness, as in such a situation it would
be more appropriate to study the direct production of bilepton pairs
rather than their virtual effects. From those Figures we can see
that the $\Delta \kappa_V$ and $\Delta Q_V$ signs are reversed, a
situation also observed in the SM model. For larger values of $m_Y$
the form factors increase with the energy, although both of them
approach asymptotically to zero for very large $|Q|$ after reaching
an extremum. This situation is similar to what is observed in the SM
after the PT is implemented \cite{PTSM1,PTSM2}. From Fig.
\ref{DeltaK}, we can see that $\Delta \kappa_V$ ranges between
$10^{-4}$ and $10^{-5}$ for a relatively light bilepton with a mass
in the range $2m_W<m_Y<8m_W$. These values are of the same order of
magnitude than the SM contribution. As far as $\Delta Q_V$ is
concerned, it ranges from $10^{-4}$ to $10^{-5}$ in the same $m_Y$
range, which means that this form factor has essentially the same
behavior than the SM contribution. Also, we can see that a more
heavy bilepton, with a mass in the range $8m_W<m_Y<12m_W$, yields
both $\Delta \kappa_V$ and $\Delta Q_V$ at the $10^{-6}$ level, and
they increase smoothly when the energy increases. This means that
for very heavy bileptons, $\Delta \kappa_V$ and $\Delta Q_V$ are one
order of magnitude smaller than the respective SM radiative
correction. Notice also that the inequality $|\Delta Q_V|>|\Delta
\kappa_V|$ always holds, which is opposite to what is observed in
the SM. This apparent contradiction stems from the fact that, in our
case, the new physics effects are of decoupling nature: the form
factors vanish in the limit of a very large mass $m_Y$. Indeed,
$\Delta Q_V$ is always of decoupling nature since it arises from a
nonrenormalizable dimension-six operator. In contrast, $\Delta
\kappa_V$ can be sensitive to nondecoupling effects since it is
associated with a renormalizable Lorentz structure of dimension
four. Its nondecoupling nature is well known from the SM
\cite{Bardin,TLVSMPT} and some of its extensions
\cite{STATICPW,TLVMSSM}.

Finally, we would like to comment our results within the context of
the effective Lagrangian framework. In this scheme, the electroweak
Lagrangian is extended with nonrenormalizable operators of dimension
higher than four respecting the $SU_L(2)\times U_Y(1)$ symmetry. In
particular, anomalous contributions to the $\Delta \kappa_V$ and
$\Delta Q_V$ form factors are induced by the following
$SU_L(2)\times U_Y(1)$-invariant dimension-six operators
\begin{eqnarray}
{\cal O}_{WB}&=& \frac{\alpha_{WB}}{\Lambda^2}(\phi^\dag {\bf
W}_{\mu \nu}B^{\mu \nu}\phi), \\
{\cal
O}_{W}&=&\frac{\alpha_{W}}{\Lambda^2}\frac{\epsilon_{ijk}}{3!}W^{i\mu}_\lambda
W^{j\lambda}_\nu W^{k\nu}_\mu,
\end{eqnarray}
where $\phi$ is the SM Higgs doublet and ${\bf W_\mu}=W^i_\mu
\sigma^i/2$, with $\sigma^i$ the Pauli matrices. The $\alpha$
constants, which parametrize the details of the underlying physics,
could be determined once the fundamental theory is known. In
addition, $\Lambda$ is the new physics scale. It turns out that the
${\cal O}_{WB}$ and ${\cal O}_{W}$ operators induce contributions to
$\Delta \kappa_V$ and $\Delta Q_V$, respectively. Also, it has been
shown that these operators can only be induced by the fundamental
theory at one-loop or higher orders \cite{AEW1}. Assuming that these
operators are induced at the one-loop level in the full theory, the
associated $\alpha$ constant must contain a factor of $1/16\pi^2$
together with an additional coefficient $g$ or $g'$ for each gauge
field. Taking the  bilepton mass  as the new physics scale, it is
natural to assume
\begin{eqnarray}
\Delta \kappa_V&\sim&
\frac{gg'}{16\pi^2}\Big(\frac{m_W}{m_Y}\Big)^2f(m_Y,m_W),\\
\Delta Q_V&\sim&
\frac{g^2}{16\pi^2}\Big(\frac{m_W}{m_Y}\Big)^2g(m_Y,m_W),
\end{eqnarray}
where $f(m_Y,m_W)$ and $g(m_Y,m_W)$ stand for the dimensionless loop
functions, whose structure depends on the details of the underlying
physics. Since the new physics effects are of decoupling nature, the
loop functions $f(m_Y,m_W)$ and $g(m_Y,m_W)$ are expected to be of
order $O(1)$ at most. Under this assumption, a straightforward
evaluation shows that $\Delta \kappa_V$ goes from $2.2\times
10^{-5}$ to $0.97\times 10^{-5}$ for $8m_W<m_Y<12m_W$, whereas
$\Delta Q_V$ ranges from $4.1\times 10^{-5}$ to $1.8\times 10^{-5}$.
This simple qualitative discussion shows that our results are in
agrement with those expected in a decoupling scenario of physics
beyond the Fermi scale.

\section{Conclusions}
\label{con} There is a plenty of good reasons to expect the
appearance of new physics beyond the SM, but it still remains a
mystery how and where this class of effects would show up. Any new
particles would arise by direct production if there is enough energy
available, or through their virtual effects on some observable. The
last scenario seems to be the most promising if the new particles
have masses much larger than the Fermi scale. In this case, high
precision measurements are needed in order to detect any deviation
from the SM predictions. We have examined this possibility via the
radiative corrections to the $WW\gamma$ and $WWZ$ vertices, which
would play a special role at NLC experiments. The one-loop
contribution to these couplings from the new gauge bosons predicted
by the minimal $331$ model was studied in a $SU_L(2)\times
U_Y(1)$-invariant way by introducing a nonlinear quantization
method. This scheme, even though conventional in the sense that it
is based on BRST symmetry, enables one to assess the new physics
effects predicted by the 331 model on the SM Green functions through
a quantum action that is invariant under the electroweak group.
Special emphasis was put on discussing the similarities between our
quantization method and the BFM. The main ingredient shared by both
methods is that they allow one to construct gauge-invariant quantum
actions. It is worth emphasizing however that while the BFM can be
used at all energies, as gauge-invariance is preserved with respect
to the gauge group of the full theory, the one presented here is
only appropriate to study heavy physics effects on low-energy (SM)
Green functions. In the latter case the complete quantum action is
only invariant under a subgroup of the theory. The $SU_L(2)\times
U_Y(1)$ invariance of the loop amplitudes associated with the
$WW\gamma$ and $WWZ$ vertices was showed and special emphasis was
put on the inherent simplicity of the calculation. Our results show
that for a relative light bilepton, with mass in the range
$2m_W<m_Y<6m_W$, both $\Delta \kappa_V$ and $\Delta Q_V$ take values
within the range of the SM contribution. On the other hand, for a
more heavy bilepton, with mass in the range $8m_W<m_Y<12m_W$, the
form factors remain essentially uniform and are of the order of
$10^{-6}$. It means that they are, respectively, two and one order
of magnitude smaller than their SM counterpart. It was also shown
that our results are in agreement with the expectations arising from
a decoupling scenario of new physics, as argued in the light of the
effective Lagrangian approach. Our results suggest that it would be
necessary a high experimental sensitivity in order to detect the
virtual effects of new massive gauge bosons because it is hard that
any heavy excitations arising from a renormalizable full theory
could be much larger than the SM radiative corrections.

\acknowledgments{We acknowledge support from SNI and Conacyt
(M\'exico). The work of JMD was supported by Conacyt under grant
U44515-F.}

\end{document}